\documentclass[aps,pre,twocolumn,superscriptaddress,floatfix,10pt]{revtex4-1}
\usepackage{graphicx}
\usepackage{dcolumn}
\usepackage{bm}
\usepackage{latexsym}
\usepackage{footmisc}
\usepackage{tikz}
\usepackage{blindtext}
\usetikzlibrary{arrows,shapes}
\def\beq{\begin{equation}}
\def\eeq{\end{equation}}
\def\bea{\begin{eqnarray}}
\def\eea{\end{eqnarray}}

\def\bc{\begin{center}}
\def\ec{\end{center}}

\begin{document}
\title[Sufficient conditions for the additivity of stall forces]{Sufficient conditions for the additivity of stall forces
  generated by multiple filaments or motors}
\author{Tripti Bameta}
\thanks{These two authors contributed equally}
\affiliation{UM-DAE Center for Excellence in Basic Sciences, University of Mumbai, Vidhyanagari
Campus, Mumbai-400098, India.}
\author{Dipjyoti Das}
\thanks{These two authors contributed equally}
\affiliation{Department of Physics, Indian Institute of Technology,
Bombay, Powai, Mumbai-400 076, India}
\author{Dibyendu Das}
\affiliation{Department of Physics, Indian Institute of Technology,
Bombay, Powai, Mumbai-400 076, India}
\author{Ranjith Padinhateeri}
\affiliation{Department of  Biosciences and Bioengineering, Indian Institute of Technology Bombay, Powai, Mumbai-400 076, India}
\author{Mandar M. Inamdar}
\affiliation{Department of Civil Engineering, Indian Institute of Technology, Bombay, Powai, Mumbai-400 076, India}

\begin{abstract}
Molecular motors and cytoskeletal filaments work collectively most of the time under opposing forces.
This opposing force may be due to cargo carried by motors or resistance coming from the cell 
membrane pressing against the cytoskeletal filaments. Some recent studies have shown 
that the collective maximum force (stall force) 
generated by multiple  cytoskeletal filaments or molecular motors may not always be just
a simple sum of the stall forces of the individual filaments or motors. To understand this excess or deficit in the collective
force, we study a broad class of models of both cytoskeletal filaments and molecular motors.
We argue that the stall force generated by a group of filaments or motors is additive, that is, the stall force of $N$
number of filaments (motors) is $N$ times the stall force of one filament (motor), when the system is in equilibrium at stall. 
Conversely, we show that this additive property typically does not hold true when the system is not at equilibrium at stall.
 We thus present a novel and unified understanding of the existing models exhibiting such non-addivity, 
and generalise our arguments by developing new models that demonstrate this phenomena.
We also propose a quantity similar to thermodynamic efficiency to easily predict  
this deviation from stall-force additivity for filament and motor collectives. 
\end{abstract}

\maketitle
%
%
%
%
%

\section{Introduction}
Molecular motors (such as kinesin, dynein and myosin)  and cytoskeletal filaments  (such as actin and microtubule) are abundantly present in  eukaryotic cells and are responsible for important cellular functions. 
Cytoskeletal filaments give structural stability to the cell and act as tracks along 
which molecular motors move and facilitate intra-cellular transportation~\cite{Lodish2007,nelsonbook,Alberts2002,Hirokawa2010610,Vale2003467,crossNatReav2014}.
Many researchers have studied the dynamics of single filaments and single motors
in great detail  using experimental and theoretical tools~\cite{bustamante2001acr,molloy1995nature,fisher1999,kolomeisky2007,kolomeisky2013jp,dcPR2013,clancy2011nsmb,tripti2013,veigel2011nrmcb,Ranjith2009,Desai_Mitchison_MT:97,Mcintosh1998,hill:84-2,schnitzer2000nat}. 
However, inside a cell, molecular motors or cytoskeletal filaments work
collectively most of the time to perform their biological tasks.
For example, the dynamics of multiple actin filaments is responsible for 
 lamellipodial protrusions during crawling of cells~\cite{rafelski2004-73,schaus_performance_2008}.
Similarly, microtubules work collectively to bring about
segregation of chromosomes during mitosis~\cite{KlineSmith2004317}. Many dynein 
motors attach to molecular cargo and generate force for cellular
transportation~\cite{efremov_delineating_2014,Mehta1999Nature}, while myosin motors primarily work together to generate forces in stress fibres and muscle tissues. 
Hence, study of such systems,  which are involved in a wide range of biological
processes,  requires understanding of the generation of collective force by multiple cytoskeletal filaments or
motors~\cite{doorn2000,levi_organelle_2006,lacostenjp2011, krawczykepl2011,Kierfeld:2013,dd2014,joanny2006prl,CasademuntPRL2009, ambarish:2008,jacprost2008BJ,demoulin_power_2014,Koster23122003,Leduc07122004}. 

To understand collective force generation by polymerizing  biofilaments, researchers typically 
resort to {\it in vitro} experiments in which biofilaments are polymerised either in the form of bundles or branched sheets against a barrier, 
which resists their  motion by producing reaction forces~\cite{Dogterom31101997,janson2004prl,Kovar12102004,Laan01072008,Brangbour2011plos,demoulin_power_2014}. 
One important focus of these studies is a quantity called the {\it stall force}, which is defined as the resisting force  at 
which the mean growth velocity of the collection of filaments is zero, and is the maximum pushing force generated by these filaments. 

A number of experiments on force generation by an assembly of biofilaments have been reported in 
the literature~\cite{Dogterom31101997,McGrath2003329,Janson23062003,Kovar12102004,Marcy20042004, Prass11092006,Parekh2005NCB,Footer2007,Laan01072008,Brangbour2011plos,demoulin_power_2014}. In view of the overall content and objective of our paper we will focus only on the experiments related to single and parallel bundles of growing  biofilaments. Experiments on single actin-filament 
are relatively rare due to experimental limitations created by the easy buckling tendency of a single actin filament. To the best of our knowledge, only one set of experiments has reported stall force for a single growing actin filament -- a value of  $\approx1$pN~\cite{Kovar12102004}. In another set of experiments on a few actin filaments growing in parallel, Footer et al.~\cite{Footer2007} reported a collective stall force of  $\approx1$~pN, 
which is very similar to the load required to stall a single filament. In this experiment, the interpretation of filament cooperativity while pushing together against the barrier is, however, complicated 
by the fact that individual actin filaments may buckle before reaching their stall force, and hence, collectively the bundle may be unable to reach its full potential for force generation. 
On the other hand, in similar experiments performed on microtubules, researchers have found that 
the growth velocity of the filament growing against an immobile barrier decreases with increased resistance -- from $1.2 \mu$m per minute
at zero load to $0.2 \mu$m per minute at 4pN to 5pN force, which is the putative stall force in this case~\cite{Dogterom31101997}. In another experiment based on microtubules, using  optical tweezers,
Laan et al.~\cite{Laan01072008} report that  stall forces of $2.7$pN, $5.5$pN and $8.1$pN are produced by groups of filaments. They interpreted this as containing one, two and three filaments,  respectively. Hence, they concluded that stall force scales linearly with the number of filaments though there could be possible ambiguity in directly counting the number of filaments in the group. 

On the theoretical and computational front, there are a number of very detailed models for understanding the force-velocity dynamics of a single biofilament~\cite{Peskin_Oster:93,mogilner1999,kolomeisky2001bj,zhang2011jbc}
, as well as a few that try to model the dynamics of multiple biofilaments~\cite{doorn2000,kolomeiskyjcp2004,lacostenjp2011,krawczykepl2011, Kierfeld:2013,wang_load_2014,dd2014, dd2014-2}. Some of these theoretical 
studies have demonstrated that the stall force of multiple, non-interacting filaments without ATP/GTP dynamics,  scales linearly with the number of 
filaments~\cite{doorn2000,lacostenjp2011,Kierfeld:2013, krawczykepl2011, kolomeiskyjcp2004}.  In contrast, a few other
recent papers quite clearly report that inclusion of ATP/GTP hydrolysis can lead to either enhanced or reduced 
cooperativity in the maximum force generated  by multiple growing filaments; the stall forces need 
not always scale with the number of filaments~\cite{dd2014,dd2014-2, kolomeisky2015,aparna2015conf}. In other words, the stall forces of individual filaments are
 non-additive in general, that is, the collective stall force produced by $N$ number of filaments (denoted by $f_s^{(N)}$) is not just a simple sum of individual stall forces of single filaments (i.e.,  $f_s^{(N)} \neq Nf_s^{(1)}$). It is quite intriguing as to why tweaking an internal variable (ATP/GTP) for an individual filament without actually changing any external interaction between the filaments can lead to this drastic change in their collective generation of force. 

In studies similar to those on the biofilaments, the stall force for motors is defined by the 
resisting force at which the average velocity of the motors is zero. 
The experiments on molecular motor force generation mostly involve micro-sized  dielectric particles and optical tweezers, by which a 
resisting force is applied on the moving motors, in order to measure their velocity response as a function of the resisting force~\cite{SVOBODA1994773,Finer1994113,roopmallik2004,Takagi20061295,Wang1998902,visscher1999nat,Mallik20052075,Jamison20102967,fallesen2011,Blehm26022013,Hendricks06112012,Rai2013172}. We briefly describe in the following text, a few examples that are relevant for our current study. 
Single molecule study of kinesin shows that kinesin 
is a strong molecular motor and generates maximum force of magnitude $\approx$ 5--8pN
~\cite{SVOBODA1994773,visscher1999nat}, whereas the stall force of certain variants of dynein is 
measured to be a relatively lower value of $\approx1$pN ~\cite{Mallik20052075}.  Dyneins in a group are, however, reported to generate force collectively, something that is missing in kinesin, mainly because of its lack of processivity under larger forces~\cite{Mallik20052075}. 
A single headed variant of the kinesin family, KIF1A,  which migrates along the microtubule in alternating states of strong attachment and incomplete detachment, produces a very small stall force ($\approx 0.1$ pN).
However, a very recent experiment of tube-pulling assay on KIF1A motors 
has demonstrated extremely strong cooperative force generation in these motors -- $\approx$ 10--15 
single headed KIF1A motors could indeed pull out tubes from 
giant unilamellar vesicles, which require a
 force around two orders of magnitude larger than the arithmetic sum of the individual stall forces~\cite{casademunt2015nature}.

There are a few broad classes of models present in the literature for understanding the force-velocity relation of both single molecular motors~\cite{julicher1997,kolomeisky2007, dill2016} 
as well as for a group of  molecular motors~\cite{kolomeisky2015sm, klumpp:2005, lipowsky:2008, Ambarish2010phybio}. The multiple molecular motor models describe a variety of different biophysical scenarios such as motors elastically coupled to each other, motors elastically coupled to the cargo, tug of war between motors walking in opposite directions and self-exclusion interactions between motors pulling on a membrane tether, for processive as well as non-processive motors~\cite{Ambarish2010phybio, joanny2006prl, klumpp:2005, lipowsky:2008, oriola2013}. Specifically, Camp\`as {\it et al.} \cite{joanny2006prl} have shown theoretically 
that the collective stall force for multiple motors are non-additive ($f_s^{(N)} \neq N f_s^{(1)}$) in the presence of
attractive or repulsive interactions and can be manyfold larger. However, in the absence of such
interactions, the stall forces in this model are simply additive.
Also, in a series of recent papers, Casademunt and co-workers
\cite{CasademuntPRL2009, oriola2013, oriola2014} using  a `two-state' 
model \cite{julicher1997} for multiple interacting motors, have demonstrated that the motors 
can produce highly enhanced cooperativity in stall force generation, and, in particular, demonstrated this phenomenon for KIF1A motor~\cite{casademunt2015nature}.

 From the above discussion, it is quite clear that the collective force generation by motors and 
 biofilaments can indeed exhibit or has the potential to exhibit highly diverse behaviour. 
As noted earlier, some studies have shown enhanced/reduced cooperativity in the collective stall 
force generation, while other studies hint towards additivity of stall 
forces. Hence,  it would be both interesting and important to understand 
the conditions under which stall forces become simply additive, and 
consequently, get better insight into the circumstances under which 
the simple additivity is lost. In this paper, we develop a general 
theoretical framework to understand how enhanced/reduced cooperativity 
in collective force generation can appear in systems of multiple 
cytoskeletal filaments or motors. We investigate this question by 
studying various models for these systems. From our case studies 
we conclude, quite generally, that the violation of stall force 
additivity  stems from the violation of the condition of 
{\it detailed balance}, that is, departure from thermodynamic 
\emph{equilibrium}. On the other hand, stall forces are additive when 
the system of filaments or motors is in equilibrium at stall. 
The main contribution of this paper is to recognize the hitherto invisible 
thread of thermodynamic equilibrium linking the phenomenon of stall 
force additivity across a variety of models.

\section{Collective stall force for multiple cytoskeletal filaments and motors: stall forces are additive for equilibrium dynamics}
\label{sec:equil}

\begin{figure}[h]
  \includegraphics[width=1.\columnwidth]{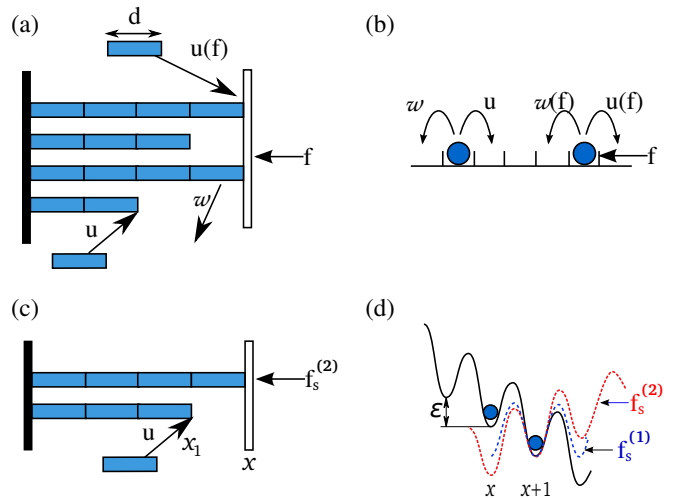}
  \caption{Biased random walk model for filaments and motors:(a) Multiple ($N>1$) filaments against a wall (see \cite{lacostenjp2011} for a detailed study of this model). Polymerization and depolymerization rates are $u$ and $w$, when the filaments are away from wall (hence force-free). Note that, when more than one filament touches the wall simultaneously, the depolymerization rate becomes independent of the force(${w}$), as a single depolymerization event cannot cause any wall movement. (b) Multiple motors moving on a track. Note that only the leading motor bears the force $f$, while the trailing motors are experiencing force through hard core interaction. (c) Two filaments
    at the collective stall force $f_s^{(2)}$ undergoing simple
    processes of polymerization and depolymerization. (d) Two motors  diffusing on a
    tilted free-energy landscape. The force is acting only on the leading motor, and the energy landscape 
    of the leading motor changes with the application of force. Due to hard-core interaction, a larger force($> f_s^{(1)}$) 
    is required to stall the system of two motors. It is like moving uphill, with the assistance of a trailing motor, which rectifies the average backward motion of the leader~\cite{Astumian917}.}
  \label{fig:1}
\end{figure}

Inspired by the growth of cytoskeletal filaments against cell
membranes, theorists have studied various models of filament dynamics
against a constant applied load.
\cite{doorn2000,lacostenjp2011,dd2014,Mogilner2003,hill:81,Peskin_Oster:93}.
Similar to the filaments, motors also work against load and
have been modelled in several studies
\cite{joanny2006prl,CasademuntPRL2009,
  ambarish:2008,klumpp:2005,lipowsky:2008}. Many of these filament and motor models are mathematically 
  very similar and can be described by  a common framework  of biased random walk (see Fig.\ref{fig:1}(a) and Fig.\ref{fig:1}(b)). We consider a collection of rigid filaments nucleating from a fixed wall at left, while a resisting force $f$ is applied upon them via a moving wall at right.
Each filament grows and shrinks by stochastic addition and removal of
subunits of size $d$ (see Fig. \ref{fig:1}(a)). Similarly, each motor 
takes a single forward or backward step on a fixed one-dimensional track 
(such as a track of microtubule or actin) of lattice constant
$d$ (see Fig. \ref{fig:1}(b)).  However, there is a key difference 
between the systems of filaments and motors: the motors cannot overtake 
each other, maintaining their sequence on the lattice, and consequently 
the leading motor alone bears  the  force $f$ (Fig. \ref{fig:1}(b)). 
The motors further obey  `mutual exclusion', that is, each lattice site can 
be occupied by one motor only when the site is empty.  On the contrary, the 
filament-tips do overtake each other since the filaments grow parallelly, 
and, therefore, any of the filaments can experience the applied force $f$ 
(Fig. \ref{fig:1}(a)). In these models, we measure the
forces in the units of $k_B T/d$, where $k_B$ is the Boltzmann constant, 
$T$ is the temperature, and $d$ is the subunit size (or lattice size). 
We take $k_B T/d =1$ without losing generality. In the absence of any force, 
we denote the
polymerization rate (forward-hopping rate) and
depolymerization rate (backward-hopping rate) for filaments (motors) as
$u$ and $w$ respectively. In the presence of a resisting force $f$, the polymerization
rates (forward-hopping rate) decreases, and the depolymerization rate
(backward-hopping rate) increases according to the following rules:
$u(f)=u e^{- f \delta}$ and ${w}(f)={w} e^{ f(1-\delta)}$. Here, the
parameter $\delta \in [0,1]$ is commonly known as force distribution
factor \cite{lacostenjp2011} [cite for motors also].

In the context of the models mentioned earlier, we can argue that the system is in equilibrium 
at stall. By definition, at the stall force, the mean velocity of the 
system is zero which implies that the overall monomer flux, in and out of the filament 
assembly, is zero. Since the monomer flux is zero, it is logical to believe that the system 
is in thermodynamic equilibrium at stall. Strictly speaking, the system is in 
equilibrium at stall, only if it is bounded in length. Unbounded systems will be freely diffusive at stall,
which is clearly a non-equilibrium phenomenon.
In fact, our class of systems (filament or motor collectives) have finite sizes  for all practical 
purposes. In the case of motors, they can be thought to diffuse on a tilted energy 
landscape (see Fig.\ref{fig:1}(d)), which results in the biased random walk. This bias can be created by chemical 
potential that is linked with the ATP hydrolysis~\cite{fisher2001}, which is irreversible and hence inherently a non-equilibrium 
process. However, in the context of these models, the role of the ATP chemical potential is just to create a tilt in the  energy landscape. Thus, as far as the translational movement of the motors against an applied force 
is concerned, we can conceptually 
think of the system to be in thermodynamic equilibrium at stall with respect to the translational degree of freedom -- the chemical 
coordinate is simply orthogonal to the translation coordinate. This seems quite analogous to the case of non-interacting active Brownian particles, which exert pressure on the confining walls similar to an ideal gas in equilibrium, but with a renormalised temperature due to the free energy consuming activity~\cite{solon2015}. Nevertheless, these arguments cannot be claimed to be true for every 
model for motors or filaments. In fact, in the subsequent sections, we show that for most models, the arguments advanced here break 
down and the systems are not in equilibrium at stall in general, except for certain choices of the parameters.

Based on these arguments, if one assumes that the system is in equilibrium at stall, the  tools of equilibrium statistical 
mechanics can be applied to calculate the stall forces. Here we show
that the stall forces are additive for multiple filaments in the 
simplest model, where only polymerization and depolymerization
processes are involved (see Fig.\ref{fig:1}(c)).  We first consider the dynamics of a single filament under 
the stall force $f=f_s^{(1)}$ applied via the right wall. Let the wall position be $x$ in terms of the subunit size (see Fig.\ref{fig:1}(c)). 
Using equilibrium statistical mechanics we can write the probability distribution of the wall-position $P(x)$ as below
\bea P(x)=\frac{1}{Z}e ^{-\beta
  f_s^{(1)} x} e^{\beta \epsilon x} =\frac{1}{Z} e^{-\beta(f_s^{(1)} -
  \epsilon)x},
\label{Px-1fila}
\eea
where $Z$ is the partition function, and $\epsilon = \ln(u/w)$ is the free energy
gained per subunit through polymerization. Note that the term $e
^{-\beta f_s^{(1)} x}$ appears as we have a Gibbs ensemble in
$1$-dimension with fixed external compressive force
($f_s^{(1)}$). However, this distribution $P(x)$ should be independent of $x$,
as the stall condition can be reached at any position $x$.  As per Eq. \ref{Px-1fila}, this
implies that $f_s^{(1)}=\epsilon = \ln(u/w)$. This same expression has been obtained in earlier theories from a detailed 
kinetic calculation~\cite{lacostenjp2011}.  Next, we consider
a two-filament system subjected to their stall force $f_s^{(2)}$ as shown in Fig.\ref{fig:1}(c). Let
the tip-position of the trailing filament be $x_1$, which is between
$0$ and the wall position $x$.  The probability distribution of the
wall-position, if the system is in thermodynamic equilibrium, is: 
\bea
&&P(x)=\frac{1}{Z}e ^{-\beta f_s^{(2)} x} e^{\beta \epsilon x} ( 2 \sum_{x_1=0}^{x} e^{\beta \epsilon x_1} -  e^{\beta \epsilon x}) \nonumber \\
&& \sim e^{-\beta(f_s^{(2)} - 2\epsilon)x}, ~~~\mbox{for large} ~~ x
\label{Px-2fila}
\eea 
A factor of $2$ appears on the right hand side, since there
could be two equally likely situations -- either the top-filament or
the bottom-filament can be the leader (see Fig. \ref{fig:1}(c)). Again, $P(x)$ is expected to be 
independent of $x$, implying that
$f_s^{(2)} = 2 \epsilon = 2 f_s^{(1)}$. This argument can be easily
extended to $N>2$, and thus $f_s^{(N)} = N f_s^{(1)}$ for this simple model.  
 Some arguments, based on detailed balance criterion, have also been given in Refs.~\cite{doorn2000} 
and \cite{krawczykepl2011} to show similar result of $f_s^{(N)}\propto N$ for their respective 
models on cytoskeletal filaments. However, we demonstrate that a simple calculation based 
on elementary statistical mechanics, for essentially kinetic processes, leads to the same conclusion.

We now develop similar arguments to show the additivity of stall forces
for multiple motors. The forward and backward hopping processes for motors
can be viewed as random walks on a tilted free-energy landscape (see
Fig.\ref{fig:1}(d)).  In this case, $x$ and $x_1$ should be
interpreted as the positions of the leading and the trailing motors respectively, for a two-motor system. 
The free energy `released' per unit step by going downhill on the
free-energy landscape is $\epsilon$ (equivalent to the polymerization
energy). For a single motor under stall force, we can write the same
equation as before  (Eq. \ref{Px-1fila}) for the probability distribution of the leading
motor's position, $P(x)$, by recognizing the system to be in equilibrium
at stall.  However, for two motors at stall, the probability distribution
of the leader's position is somewhat different from the previous case of
the filaments, since motors cannot overtake each other. The distribution of the leader's position  is  
\bea
P(x)&=&\frac{1}{Z}e ^{-\beta f_s^{(2)} x} e^{\beta \epsilon x} (  \sum_{x_1=0}^{x-1} e^{\beta \epsilon x_1}) \nonumber \\
&& \sim e^{-\beta(f_s^{(2)} - 2\epsilon)x}, ~\mbox{for large} ~ x. 
\eea
Again, using the argument that $P(x)$ should be independent of $x$, we
get back the force additivity : $f_s^{(2)} = 2 \epsilon = 2
f_s^{(1)}$.

In summary, we show that the stall forces are additive for the simple models considered here. 
This demonstration hinges on the recognition that the systems are in equilibrium at stall, which may not be
true in many biological situations. In fact, we show in the next sections that there are many 
classes of models for which the equilibrium description is simply not feasible, and consequently more complex models have interesting implications. 
\section{Stall forces are non-additive for biologically relevant non-equilibrium models}
\label{sec:3}

In this section, we present several case studies to show that the
force inequality ($f_s^{(N)} \neq N f_s^{(1)}$) is true in general for
stall dynamics departing from equilibrium -- however, for certain combinations of kinetic rates 
the relationship $f_s^{(N)} = N f_s^{(1)}$ can be retrieved. We begin by analyzing various models of cytoskeletal filaments.

\subsection{A random hydrolysis model for cytoskeletal filaments}
\label{sec:random}

\begin{figure}[t]
  \includegraphics[width=1.\columnwidth]{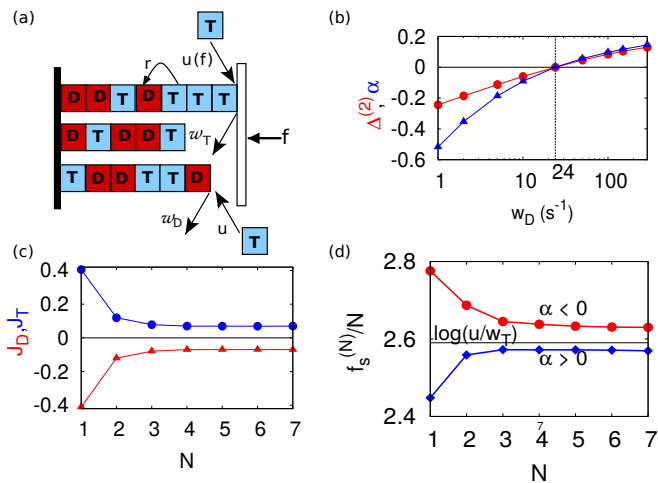}
  \centering
  \caption{ Random hydrolysis model with three filaments, where
    individual monomer switches from T to D unidirectionally and
    randomly. Different processes are shown by arrows. $k_0=3.2 ~\mu {\rm M}^{-1} s^{-1}$, $c=100\mu$M, $ w_T=24 s^{-1}$, and $r=0.2 s^{-1}$. (a) $\Delta^{(2)}$ (red curve) and   $\alpha$ (blue curve) versus $w_D$. Note that both $\alpha$ and $\Delta^{(2)}$ are zero only at $w_T=w_D$ and they are highly correlated in sign. (b) The fluxes per filament (defined in the text), at stall, for T and D monomers as a function of the filament number $N$ for $w_D=290 s^{-1}~(\alpha>0)$. (c) The collective stall force per filament ($f_s^{(N)}/N$) against $N$, the number of filaments. For $\alpha > 0$, $w_D=290 s^{-1}$ and for $\alpha <0$ , $w_D=5 s^{-1}$}
  \label{fig:2}
\end{figure}

In cytoskeletal filaments (such as, microtubules and actin filaments),
subunits  are typically bound to ATP/GTP molecules. When the subunits are connected to the filaments, the ATP/GTP molecules release phosphate and convert to ADP/GDP in a  process known as ATP/GTP hydrolysis
\cite{howardbook,Alberts2002}. The ADP/GDP-bound monomers have much
higher depolymerization rates compared to ATP/GTP-bond monomers~\cite{Pollard1986,Desai_Mitchison_MT:97}. Due
to this heterogeneity the cytoskeletal filaments exhibit interesting properties such as `dynamic instability'~\cite{PhysRevLett.70.1347}. 
The dynamics of the cytoskeletal filaments have
been theoretically studied by many researchers using the ``random
hydrolysis'' model
\cite{vavylonis:05,Ranjith2010,Ranjith2012,sumedha,Antal-etal-PRE:07}. Here, 
we focus on the simplified model of this process as discussed in Refs.~\cite{dd2014,Ranjith2012}, 
where  we neglect complexities related to biofilaments such as structural details, mechanical flexibility,  
possible interplay between mechanical forces and hydrolysis events and the possible multi-stage nature of 
the hydrolysis process~\cite{howardbook,Desai_Mitchison_MT:97,Korn638,vavylonis:05,10.1371/journal.pbio.1001161}. As shown in Fig. \ref{fig:2}(a), we consider multiple filaments undergoing random
hydrolysis and growing against a wall held by a constant opposing force
$f$. In the model, each monomer can be in two states: T
(ATP/GTP-bound), and D (ADP/GDP-bound). Only T monomers bind to the
filaments with a rate $u(f) = u {\rm e}^{f}$ (next to the wall) or $u
$ (away from the wall). The rate $u$ is proportional to the
concentration ($c$) of T monomers and is defined as $u=k_0
c$. The depolymerization occurs with a rate $w_T$ if the tip-monomer
is T, and $w_D$ if it is D. For simplicity we assume that there is no
force dependence on the off-rates $w_T$ and $w_D$ (i.e., $\delta =1$). Hydrolysis (T to D conversion) happens randomly on any T subunit in space with a rate $r$. Note that the conversion T$\rightarrow$D is
irreversible, as it is not balanced by a reverse conversion, which
makes the dynamics inherently non-equilibrium. The exact
analytical result for the stall force $f_s^{(N)}$ is not known for
such a detailed model. Instead we numerically
simulate the model using the Gillespie algorithm~\cite{Gillespiejpc1977} (see Appendix A) with experimentally known rates \cite{howardbook,Pollard1986,Desai_Mitchison_MT:97} for
microtubules and actin filaments.

Before proceeding to discuss the issue of stall force additivity, we  define a parameter, the ``force deviation'': 
\bea
\Delta^{(N)}&=&f_{s}^{(N)}-N~f_s^{(1)} 
\eea
 This parameter represents the excess/deficit of the collective stall force generated by $N$ filaments, as compared to $N$ times the force generated by a
single filament. So the deviation $\Delta^{(N)} \neq 0$
 implies the violation of force equality,
$f_s^{(N)}=Nf_s^{(1)}$.

As noted before, the random hydrolysis model is a non-equilibrium model by 
construction. To understand the implications 
of this non-equilibrium nature, we first
look at the energetics associated with the
polymerization/depolymerization processes.  A growing filament clearly
performs work against an applied load $f$ through polymerization. The
work done for addition of one subunit to the filament is simply $f d$, where we can
take subunit-size $d=1$ without losing any generality. At the stall
force $f=f_s^{(1)}$,
the filament delivers the maximum work $W_{\rm poly}^{\rm max} =
f_s^{(1)}$. The free energy input to the filament in order to do this work is 
provided by polymerisation and can be written as $F_{\rm poly}=\ln (u/w_T)$, per
subunit addition. Note that D monomers do not polymerize, and hence there is
no contribution in $F_{\rm poly}$ due to D monomers. Finally we define
the following quantity for a single filament
 \beq \alpha = F_{\rm
  poly} - W_{\rm poly}^{\rm max} = \ln (u/w_T) - f_s^{(1)}
\label{eq:alpha}
\eeq
The quantity $\alpha$ signifies how different the maximum work produced
per filament is, as compared to the free energy input. Hence, $\alpha$ is analogous to the
thermodynamic efficiency of the system \cite{niraj:prl-efficiency}. 
%
\begin{figure*}
  \centering
 \includegraphics[width=1.0\textwidth]{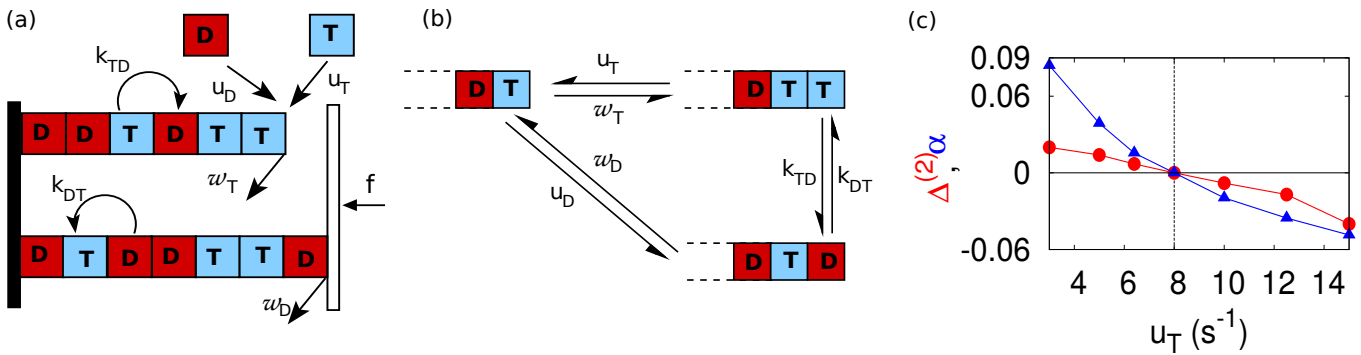}
  \caption{ (a) Schematic diagram of a generalized random hydrolysis model with two-way switching  (both T$\rightarrow$D, and D$\rightarrow$T). Different processes (shown in arrows) are discussed in the text. (b) Schematic depiction of a connected loop in the configuration space
of a single filament, within the  model. (c) Deviation ${\Delta}^{(2)}$  versus $u_{T}$ (the red curve), and $\alpha$  versus $u_{T}$ (the blue curve), for the generalized random hydrolysis model. The parameters are: $w_{T}=2 s^{-1}, k_{TD}=0.3 s^{-1}, k_{DT}=0.4 s^{-1}, u_{D}=3 s^{-1} $, and $ w_{D }=1 s^{-1}$.}
\label{fig:3}
\end{figure*}
%
In Fig.\ref{fig:2}(b) we plot the deviation $\Delta^{(2)}$ and the efficiency $\alpha$ for two
microtubules,  against the dissociation rate $w_D$ of D
monomers. Interestingly,
$\Delta^{(2)}$ and $\alpha$ are correlated in the numerical sign. This gives us a
hint that the violation of force additivity has something to do with
the imbalance between the work produced and the free-energy input
(i.e., the departure from equilibrium). To appreciate this 
interconnection between $\Delta^{(2)}$ and $\alpha$, we proceed to investigate the particle fluxes of
the T and D monomers.

We calculate the particle fluxes when the $N$-filament system is at
stall.
In the simulations~\cite{Gillespiejpc1977} we  separately keep track of the numbers of T and D
monomers binding (or unbinding) at a filament-tip in a $N$-filament
system. The fluxes are then calculated over a time
window. The flux for T monomers per filament ($J_T$) is defined as the net change of T monomer numbers at any one filament-tip, divided by the size of the
time window. Similarly, we also calculate the flux per filament for
D monomers ($J_D$). In Fig. \ref{fig:2}(c), we show these fluxes at stall. 
 Although we have $J_T+J_D =0$ (which is expected at stall), individually the fluxes per filament  (both $J_T$ and $J_D$) are non-zero signifying the
non-equilibrium nature of the dynamics. An important point to note is that, the fluxes per filament, at stall, decrease with the number of filaments, $N$,
and tend to saturate. From this observation, we are tempted to make a
hypothesis that a \emph{non-equilibrium system with a large number of filaments is 
closer to ``equilibrium'' in comparison to a single-filament system}(see Appendix  B for a crude entropy production argument for this hypothesis). 

With this hypothesis in hand, we now attempt to explain why the numerical signs of $\Delta^{(2)}$ and $\alpha$ are correlated (Fig. \ref{fig:2}(b)). We first consider the case $\alpha>0$, when a single filament performs less work than the
free-energy provided by the polymerization (see Eq. \ref{eq:alpha}). In this case, some energy is dissipated by the filament due to the internal T$\rightarrow$D
transitions. As per our hypothesis, if the two-filament system is 
closer to equilibrium  as compared to a single filament, the 
two-filament system should extract more work by increasing 
the stall force per filament. To check this, in Fig.\ref{fig:2}(d), we show the stall force per filament (i.e., the maximum work extracted per filament), $f_s^{(N)}/N$
as a function of the filament number $N$. The stall force per filament 
 indeed increases with $N$ for $\alpha>0$, and
saturates near the net free-energy input ($\ln (u/w_T)$). In other words, as the number 
of filaments increases, the collective stall force per filament gets closer to the ``equilibrium" value of $\ln(u/w_T)$. This increase in stall 
force per filament makes $f_s^{(2)}/2>f_s^{(1)}$ and in return gives positive $\Delta^{(2)}$. 
Thus, the $\alpha>0$ case correlates
with $\Delta^{(2)} >0$. Similar
arguments can be given for $\alpha<0$ (see Fig.\ref{fig:2}(d)), 
where a single filament performs more work than the energy provided by polymerization. 
To bring the system closer to equilibrium, the two-filament system decreases the stall force per filament ($f_s^{(N)}/N<f_s^{(1)}$). Hence, $\Delta^{(2)}$ is negative if $\alpha<0$.

An interesting point to note in Fig.\ref{fig:2}(b) is that both
$\Delta^{(2)}$ and $\alpha$ are zero exactly at $w_T = w_D$. 
This
shows that T$\rightarrow$D switching (hydrolysis) is necessary to
produce the phenomenon of non-additivity of stall forces.
 It is to be noted that the hydrolysis  is always a non-equilibrium process as it is
unidirectional. However, the condition $w_T = w_D$ effectively
corresponds to absence of switching, since dynamically there remains no
distinction between T and D subunits. The filaments cannot ``sense''
their distinct presence as far as the force generation is
concerned. 
Yet, the condition $w_T = w_D$ does not
imply a {\it true}  equilibrium until we set the T$\rightarrow$D
switching rate to zero. Another way to possibly achieve equilibrium at stall is to
incorporate the reverse switching (D$\rightarrow$T) and allow the polymerization of D subunits.  Although these additions 
are biologically unrealistic, we nevertheless study such a model in the next section 
to explore the relevance of  non-equilibrium dynamics for
non-additivity of the stall forces.


\subsection{A generalized random hydrolysis model for filaments}
\label{sec:rev-hydro}

We make the random hydrolysis model (discussed in section \ref{sec:random}) more general and symmetric by allowing (i) D $\to$ T
conversion, and (ii) addition of both D and T monomers to the filaments. In this model
(see Fig.\ref{fig:3}(a)) both T and D subunits can bind to a filament with
constant rates $u_T$ and $u_D$, respectively. When the filaments come in contact with 
the wall (see Fig.\ref{fig:3}(a)), the polymerization
rates decrease to $u_T(f)=u_T \rm{e}^{-f}$ and $u_D(f)=u_D
\rm{e}^{-f}$ in the presence of the force $f$ (using the load distribution factor $\delta = 1$ for simplicity). 
The depolymerization occurs with a rate $w_T$ if the
tip-monomer is T, or $w_D$ if it is D. Any randomly chosen subunit
inside a filament can convert either from T to D (with rate $k_{TD}$),
or from D to T (with rate $k_{DT}$).

Within the general version of the random hydrolysis model, we now
proceed to show that the two-way switching (T $\to$ D, and D $\to$ T)
in general produces non-equilibrium dynamics that is embodied in the violation
of the condition of {\it detailed balance} for the kinetic rates. For a single filament, as
shown in Fig. \ref{fig:3}(b), we consider a loop of dynamically
connected configurations.  The product of  clockwise and
anticlockwise rates are $u_T k_{TD} w_D$ and $u_D k_{DT} w_T$,
respectively. For the condition of detailed balance, that is, 
equilibrium, to be reached at steady state, the two products must be equal according to the {\it
  Kolmogorov's criterion} ~\cite{kolmogorov1936,book-kolmogorov} (or the Wegschieder condition~\cite{Wegsch}), which leads to
\begin{equation}
 \frac{u_{T}}{w_{T}} \frac{k_{TD}}{k_{DT}} \frac{w_{D}}{u_{D}}=1.
\label{loop:rev}
\end{equation}
If we fix the parameters $w_{T}=2 s^{-1}, k_{TD}=0.3 s^{-1},
k_{DT}=0.4 s^{-1}, u_{D}=3 s^{-1} $, and $ w_{D }=1 s^{-1}$, then we
would have $u_{T}=8 s^{-1}$ from the above equilibrium condition
(Eq. \ref{loop:rev}). Though this criterion is written in terms of force-free rates, 
it is clear that using the modified rates in the presence of the resisting force 
would not change the condition in Eq. \ref{loop:rev}. We now plot the deviation $\Delta^{(2)} =
f_s^{(2)}-2f_s^{(1)}$ versus $u_{T}$ in Fig.\ref{fig:3}(c) -- see the red curve (data from stochastic simulation). The plot quite
interestingly shows that $\Delta^{(2)}=0$ only at $u_{T}=8 s^{-1}$;
otherwise it is nonzero. This clearly indicates that the phenomenon of
non-additivity of stall forces is tied to the departure of the system from
equilibrium. This can be compared with the arguments given in
section \ref{sec:equil}, where it is shown that the filament
models involving no switching exhibit equilibrium at stall, and consequently the relationship $f_s^{(N)} = N
f_s^{(1)}$ holds without any restriction. 


We can further relate
the non-equilibrium effect of non-additivity of the force with the
imbalance between the free energy input and maximum work output (as discussed in section~\ref{sec:random}). Because, in the present model,
 the filaments can grow by adding D or T monomers, 
the free energy input to the system should depend on polymerization energies corresponding to both T and D monomers. 
Consequently, the partition function for a single filament is $e^{\epsilon_T}+e^{\epsilon_D}$ (using $k_B T = 1$), where 
$\epsilon_T$ and $\epsilon_D$ are polymerization energies provided by T and D monomers, 
respectively. Hence,  the free energy input to the filament is 
$F_{\rm poly}=\ln [e^{\epsilon_T} + e^{\epsilon_T}]=\ln [(u_{T}/w_T) +(u_{D}/w_D)]$; 
and the maximum work done by the filament is simply $f_s^{(1)}$ (using subunit size $d = 1$). 
Following previous section~\ref{sec:random}, we again define an efficiency-like parameter for the current model as below 
\beq \alpha =
F_{\rm poly} - W_{\rm poly}^{\rm max} = \ln [(u_{T}/w_T) +
(u_{D}/w_D)] - f_s^{(1)}
\label{eq:alpha-rev}
\eeq
 We see in Fig.\ref{fig:3}(c) that both $\alpha$ and
$\Delta^{(2)}$ are correlated in numerical sign, and they are non-zero every
where except for a single point where the system is in equilibrium at stall
(according to Eq. \ref{loop:rev}).  With this understanding of the
connection between the non-equilibrium dynamics and non-additivity
of stall forces, we study another model for filaments \cite{dd2014} in the next section, and show that the same ideas can be carried forward.
\begin{figure*}
\includegraphics[width=1.0\textwidth]{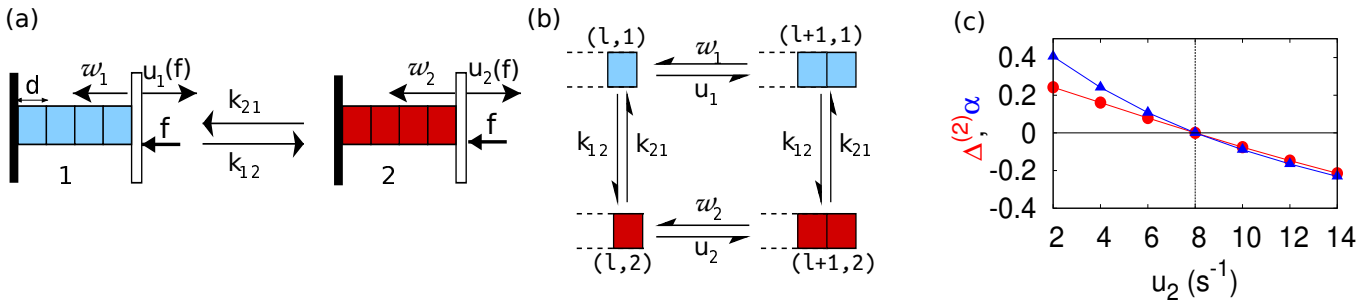}
\centering
\caption[2-state model]{(a) Schematic diagram of single-filament 2-state model with switching between  states $1$ (blue) and $2$ (red). Various processes are shown by arrows and corresponding rates, as discussed in the
text. (b) Schematic depiction of a connected loop in the configuration space
of single-filament 2-state model. The configurations are denoted by ordered pairs, whose first element is the instantaneous length and second element is the chemical state ($1$ or $2$). (c)$\Delta^{(2)}$ (red) and
    $\alpha$ (blue) as a function of $u_{2}$ for the 2-state model. Parameters are $k_{12}=0.5 ~s^{-1}$, $k_{21}=0.5~s^{-1}$, $w_{1}=0.1~s^{-1}$ $u_{1} = 1~s^{-1}$,
    $w_{2}=0.8~s^{-1}$. }
\label{fig:4}
\end{figure*}

\subsection{A two-state model for filaments }
\label{sec:toy}

In the literature~\cite{dd2014,hill:84-2}, there exists a model which incorporates the detailed process of hydrolysis in a more coarse-grained way.
In this model (Fig.\ref{fig:4}(a)),  each filament
can switch between two chemical states $1$ and $2$, with switching
rates $k_{12}$ (from $1$ to $2$), and $k_{21}$ (from $2$ to $1$). In
the states $1$ and $2$, the filament has distinct depolymerization
rates $w_1$ and $w_2$, respectively, and polymerization rates of $u_1$ and $u_2$, respectively (see Fig.\ref{fig:4}(a)).If a filament bears
the load (i.e., touches the wall), its polymerization rate is modified
to $u_i (f)=u_i {\rm e}^{-f}$ $(i=1,2)$.  For simplicity, we assume that the depolymerization rates
 are force independent.

For the above two-state model, we first explicitly show
that at stall the dynamics is non-equilibrium. As shown in Fig.\ref{fig:4}(b), we consider a loop of
connected configurations for a single filament characterized by its length and state. In this case, 
Kolmogorov's criterion reduces to 
\bea
&& u_1 k_{12} w_2 k_{21}=k_{12} u_2 k_{21} w_1 \nonumber\\
&& \Longrightarrow  \frac{u_{1}}{{w}_{1}}=\frac{u_{2}}{{w}_{2}}
\label{eq:toy}
\eea
Following the procedure
involving microscopic master equations, as described in
Ref.~\cite{dd2014}, we analytically obtain the single-filament and
two-filament stall forces ($f_s^{(1)}$, $f_s^{(2)}$) 
In Fig.\ref{fig:4}(c),
we plot $\Delta^{(2)}$ against $u_{2}$ (red curve) with fixed $w_{1} =
0.1s^{-1}$, $u_{1} = 1s^{-1}$, and $w_{2} = 0.8s^{-1}$. We find
that ${\Delta}^{(2)} \neq 0$ for all $u_{2}$, except for $u_{2} =
8s^{-1}$ (the value corresponding to the equality in Eq.
\ref{eq:toy}).  This shows that the non-additivity of
stall forces is tied to the departure from equilibrium.

It is to be noted that in the two-state model ,  $\ln (u_{i}/{w}_{i})$ is the
polymerization free energy in state $i=(1,2)$. Hence, if $P_i$ is the
probability of finding a filament in a state $i$, then at any instant
the amount of free energy that is transferred from the bath of
monomers to the filament by addition of monomers of type $1$ or $2$ is
 $F_{\rm poly} = P_1 \ln (u_{1}/{w}_{1}) + P_2 \ln
(u_{2}/{w}_{2})$. On the other hand, the maximum amount of work done
by a filament against the applied force is  $W^{\rm max} =
f_s^{(1)}$ (monomer size being $d=1$).  Therefore, as defined in the previous sections, we can again define an
efficiency parameter as:
 \beq 
\alpha = F_{\rm poly} - W^{\rm max} =[P_1 \ln (u_{1}/{ w}_{1}) + P_2 \ln (u_{2}/{w}_{2})] - f_s^{(1)},
\eeq
where, $P_1=k_{21}/(k_{12}+k_{21})$ and
$P_2=k_{12}/(k_{12}+k_{21})$. These probabilities follow from the fact that the
detailed balance relation $P_1 k_{12}=P_2 k_{21}$ holds at the steady
state for a single filament, as intuitively evident from the Fig.\ref{fig:4}(a)
(also see Ref \cite{dd2014}),
along with the normalization condition $P_1 + P_2 =1$. In
Fig.\ref{fig:4}(c), we plot $\alpha$ versus $u_{2}$ (blue
curve). We see that $\Delta^{(2)}$ is closely coupled to $\alpha$ in
numerical sign, and both are nonzero except at the point where
the equilibrium condition (Eq. \ref{eq:toy}) is satisfied.


The effect of non-additivity of stall forces is not specific to
cytoskeletal filaments. Even the system of multiple molecular motors show
such a behavior \cite{joanny2006prl,casademunt2015nature}. In the ensuing sections, we explore the connection 
between the non-additivity of the force  with the non-equilibrium dynamics in the system of motors.


\subsection{Model of interacting motors by Camp\`as et al.}
\label{sec:campas}

\begin{figure}[h]
\includegraphics[width=1.0\columnwidth]{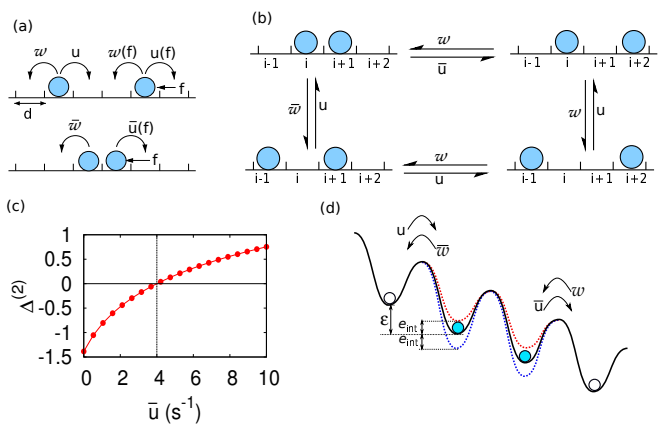}
\centering
\caption{(a) Schematic diagram of the motor model proposed by Camp\`as et al. \cite{joanny2006prl}, 
where multiple interacting motors push against a cargo with a constant force $f$ acting against their motion. 
Various processes are shown by arrows and discussed in the text. (b) Schematic depiction of a 
closed loop of dynamically connected configurations for the model shown in (a). (c)Deviation ${\Delta}^{(2)}$  versus $\bar{u}$ within the model of 
Camp\`as et al (data obtained from the exact formula given in \cite{joanny2006prl}). 
Parameters are specified in the text. We took the force distribution factor $\delta=1$.(d)Motors walking on a free-energy landscape~\cite{joanny2006prl}. The effect of nearest-neighbour interactions is shown schematically.}
\label{fig:5}
\end{figure}

A model of multiple interacting motors pushing against a load has been
recently proposed by Camp\`as et al. \cite{joanny2006prl}  . In this model (Fig.~\ref{fig:5}a), motors walk along a one-dimensional lattice
(lattice spacing $d = 1$) and move by a single step forward (rate $u$)
or backward (rate $w$). There is hard core interaction between the
motors. The leading motor alone bears the load, and hence its hopping
rates are modified to $u(f)=u {\rm e ^{-f\delta} }$, and $w(f)=w {\rm
  e ^{f (1-\delta)} }$, where
$\delta$ is the force distribution factor. The hopping rates also change 
due to nearest-neighbour interactions -- if a motor is
adjacent to another one, its forward and backward hopping rates become
$\bar{u}$ and $\bar{w}$, respectively (see Fig. \ref{fig:5}(a)).

From analytical calculations and numerical simulations, the authors
have found that the stall forces are not necessarily additive. We show here 
that the non-additivity is a manifestation of the
non-equilibrium nature of the dynamics. To show this we apply
the Kolmogorov criterion, by making a closed loop of connected
configurations as shown in Fig.\ref{fig:5}(b). By equating the
clockwise and counter-clockwise products of rates along the loop, we
have 
\bea && u. \bar{u}. w. w= \bar{w}. u. u. w\nonumber\\ 
&& \Longrightarrow \frac{u}{w}=\frac{\bar{u}}{\bar{w}}
\label{loop-campas}
\eea

Exact analytical expressions of the single-motor stall force
$f_s^{(1)} ={\rm ln}(u/w)$, and the two-motor stall force $f_s^{(2)} =
{\rm ln}[(u \bar{u}/w \bar{w}) + (u/w) - (\bar{u}/\bar{w})]$ are
derived in Ref \cite{joanny2006prl}. If we put the equality of
Eq. \ref{loop-campas} in the expression of $f_s^{(2)}$, we clearly
see that the relationship $f_s^{(2)}=2 f_s^{(1)}$ follows.  We
further plot (Fig. \ref{fig:5}(c))
$\Delta^{(2)}=f_s^{(2)}-2f_s^{(1)}$ versus $\bar{u}$, with fixed
$u=20s^{-1}$,$w=5s^{-1}$, and $\bar{w}=1s^{-1}$. In this case,
$\bar{u}=4s^{-1}$ corresponds to equilibrium dynamics
(Eq. \ref{loop-campas}), and we see that except for $\bar{u}=4s^{-1}$,
$\Delta^{(2)} \neq 0$ everywhere. This corroborates our hypothesis that the additivity of stall forces is closely linked with the
underlying equilibrium of the system, a point not recognized in Ref.~\cite{joanny2006prl}.


We can also obtain the condition of equilibrium (Eq.~\ref{loop-campas})
from a simple thermodynamic argument by considering the
hopping processes of the motors on a free-energy landscape  (see
Fig.~\ref{fig:5}(d)). Due to the nearest-neighbour
interactions, the shape of the free energy landscape gets altered when the motors
are adjacent to each other. An attractive interaction deepens the energy
wells by an amount $e_{\rm int}$ (dotted blue curve in
Fig.~\ref{fig:5}(d)), and makes it hard for the
particles to leave the position. This reduces the forward and backward
hopping rates. On the other hand, the repulsive interaction makes the potential
wells shallower (dotted red curve in Fig.~\ref{fig:5}(d)),
which makes it easy for the particles to hop forward or backward. In
this case, hopping rates increase from the original values. 
When the motors are adjacent to each other, following
Fig.~\ref{fig:5}(d), the hopping rates are given by 
\bea
\label{rate-eq-campas1}
\frac{u}{\bar{w}}&=&e^{\epsilon -e_{\rm int}}\\
\frac{\bar{u}}{w}&=&e^{\epsilon +e_{\rm int}}
\label{rate-eq-campas2}
\eea
For a motor which is  away from the other one, $e_{\rm int}=0$, and we further have
\beq
\frac{u}{w}=e^{\epsilon}
\label{rate-eq-campas3}
\eeq 
By rearranging the equations \ref{rate-eq-campas1},
\ref{rate-eq-campas2} and \ref{rate-eq-campas3}, we get back the same condition, 
$\frac{u}{w}=\frac{\bar{u}}{\bar{w}}$, as in
Eq. \ref{loop-campas}, which is a reflection of the equilibrium dynamics at stall.


\subsection{A motor model with multiple step-sizes}
\label{sec:motor-2steps}

\begin{figure*}[t]
\includegraphics[width=1\textwidth]{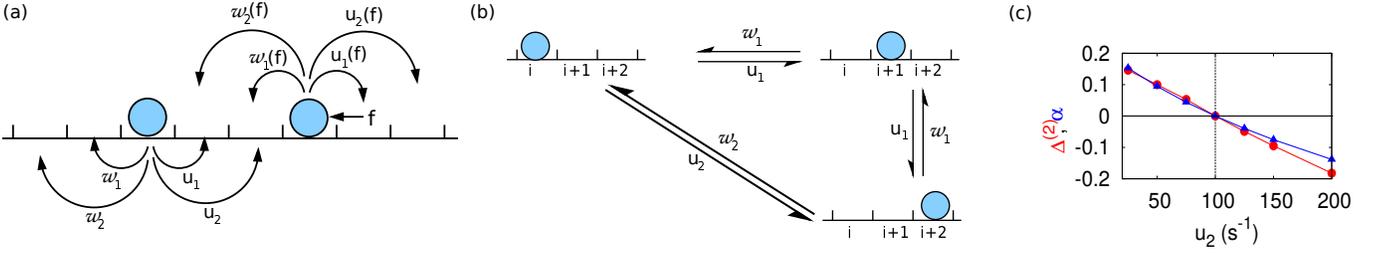}
\caption{Multiple step size motor model: (a) Schematic diagram of the motors 
moving with two distinct step sizes. The kinetic processes (shown in arrows) are 
explained in the text. (b) A closed loop of connected configurations for the model with one motor. (c) Deviation ${\Delta}^{(2)}$  versus $u_2$ (the red curve), and $\alpha$  
versus $u_2$ (the blue curve). The parameters are $u_{1}=80 s^{-1}, w_{1}=8s^{-1}$, and $w_{2}=1s^{-1}$. }
\label{fig:6}
\end{figure*}

It was found \cite{tripti2013,roopmallik2004} recently that
dynein motors on a microtubule can take multiple step sizes --
predominantly $24$-nm and $32$-nm steps. This inspired us to cast a new
model of motors walking with two distinct step sizes (see
Fig. \ref{fig:6}(a)).  A motor at a lattice site $i$ can hop to any of the sites $i+1$ (with rate $u_1$), $i+2$ (with rate $u_2$), $i-1$
(with rate $w_1$), or $i-2$ (with rate $w_2$). The leading motor
alone bears the applied force $f$ and its
forward rates are modified to $u_1(f)=u_1 {\rm e ^{-f} }$ and
$u_2(f)=u_2 {\rm e ^{- 2 f} }$, while the backward rates are assumed to be
force-independent. Unlike the model in the previous section (Section~\ref{sec:campas}) there is no explicit 
attractive or repulsive interaction between the motors.

Proceeding in a similar way as described in section
\ref{sec:rev-hydro}, we first derive the
condition for equilibrium, by considering a closed loop of connected
configurations as shown in Fig. \ref{fig:6}(b). Following
Kolmogorov criterion we get 
\bea
&& u_1^2.  w_2 = u_2.  w_1^2\nonumber\\
&& \Longrightarrow \frac{u_2}{w_2}= \left(\frac{u_1}{w_1} \right)^2
\label{loop-2-step}
\eea
 If we take $u_{1}=80 s^{-1}, w_{1}=8s^{-1}$, and $w_{2}=1s^{-1}$,
then $u_{2}=100s^{-1}$ is the value corresponding to the equilibrium
condition (see Eq. \ref{loop-2-step}). In Fig. \ref{fig:6}(c),
$\Delta^{(2)}$ versus $u_{2}$ is plotted  from our
simulation results (the red curve), where we see that indeed $\Delta^{(2)} =0$ only
for $u_{2}=100 s^{-1}$. Thus, one can again associate the force
inequality $f_s^{(N)} \neq N f_s^{(1)}$ to the violation of the
detailed balance condition.

Just like the models of filaments discussed in the previous sections (Sections \ref{sec:random},
\ref{sec:rev-hydro} and \ref{sec:toy}), we can show that the effect of force-additivity for motors is
related to the underlying energetic imbalance. The motor model with multiple step-sizes 
 looks conceptually 
similar to the generalized random hydrolysis model (see Section \ref{sec:rev-hydro}). At any instant, 
a filament can grow by adding  T or D monomers in 
the generalized random hydrolysis model; whereas
in the current model a motor can move forward (or backward) by taking steps of sizes $d$ or $2d$. 
These steps of size $d$ and $2d$ involve different amounts of work done by a single motor at stall ($f_s^{(1)}d$ and $f_s^{(1)}2d$ respectively).
By contrast,  addition of a D or T monomer to a single filament leads to the extraction of the same amount of work ($f_s^{(1)}d$). 
Hence, for  the definition of the efficiency parameter, $\alpha$, we choose to take into 
account the energy imbalance that results from  a single step of size $d$. 
For a single motor taking a step of unit lattice size, 
we can write the free-energy supplied by ATP molecules as $F=\ln
(u_1/w_1)$ and the maximum work done 
as $W^{\rm max}= f_s^{(1)}$ (considering the lattice spacing $d=1$). Thus,
we can define an efficiency-like quantity as
 \beq \alpha = F - W^{\rm
  max} = \ln (u_1/w_1) - f_s^{(1)}.
\label{eq:alpha-2step}
\eeq
We do not claim that this is a unique definition of $\alpha$ for the system. One may certainly come up 
with some other definition of $\alpha$, for example, 
\bea
\alpha =  {1 \over 2} (\ln{u_2 \over w_2} - 2 f_{s}^{(1)}) \label{eq:alpha2},
\eea
to quantify the excess/deficit of energy
supply to the system. Note that both the definitions of  
$\alpha$ (Eqs.~\ref{eq:alpha-2step} and \ref{eq:alpha2}) essentially capture
the energy imbalance per unit step. 

We plot $\alpha$ (from Eq.~\ref{eq:alpha-2step}) and $\Delta^{(2)}$ versus $u_2$ in Fig.\ref{fig:6}(c), 
and see that both $\alpha$ and $\Delta^{(2)}$ are indeed
correlated in numerical sign. Moreover, both of them are zero only when the equilibrium condition
(Eq. \ref{loop-2-step}) is satisfied. The same finding can be derived using the other definition of $\alpha$, that is, using Eq.~\ref{eq:alpha2} instead of Eq.~\ref{eq:alpha-2step} (data not shown). This further strengthens our point that non-additivity of stall forces is a manifestation of underlying
non-equilibrium stall dynamics.

\section{Discussion and conclusion}
\label{discussion}
Collective force generation by filaments or motors has been
theoretically studied by many researchers using various models in specific
contexts~\cite{doorn2000,lacostenjp2011,Kierfeld:2013,dd2014,joanny2006prl,CasademuntPRL2009,kolomeiskyjcp2004,dd2014-2,Ambarish2010phybio,Bouzat2010phybio,PhysRevE.91.022701}. However, a broad picture explaining the
cooperative effects in stall (maximum) force generation is still missing.
In this paper we have provided a theoretical framework to understand and predict the
cooperative effects in the maximum force generation by multiple motors
or filaments, for a broad class of models. It is now appreciated, at least theoretically, 
that the stall force of individual
cytoskeletal filaments or molecular motors, when they push
  together against some obstacle, is not additive in general~\cite{dd2014,joanny2006prl, kolomeisky2015}. In this paper,
we have provided several pointers to show that this non-additivity of the stall forces
($f_s^{(N)} \neq N f_s^{(1)}$) is a manifestation of the underlying
non-equilibrium nature of the dynamics at stall. 
\begin{figure}[h]
\includegraphics[width=1.0\columnwidth]{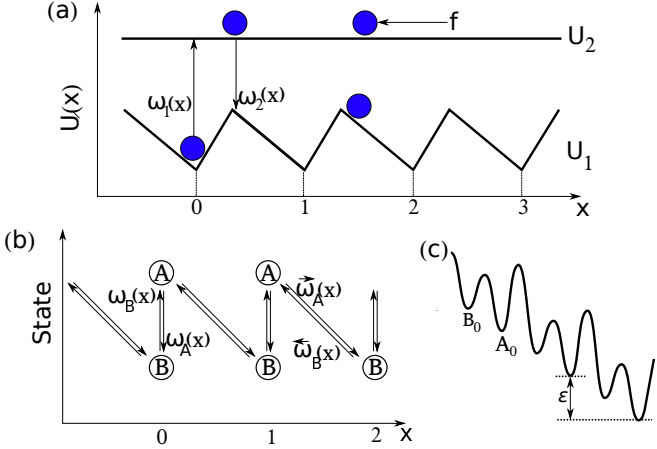}
\centering
\caption{Schematic of the (a) continuous~\cite{julicher1997} and (b) discrete two-state Brownian ratchet model~\cite{lacoste2007, lacoste2008}. 
(a) A spontaneous motion is expected when the ratio of transition 
rates $\omega_1(x) \over \omega_2(x)$ is  far from the equilibrium 
value given by the detailed balance condition.(b) Discrete version of the two-state ratchet model with effective transition rates $\protect \overrightarrow{\omega}_A(x)$, $\protect \omega_{A}(x)$, $\protect \omega_{B}(x)$ and $\protect \overleftarrow{\omega}_B(x)$.(c)Energy landscape corresponding to discrete two-state model.}
\label{fig:rachet}
\end{figure}

Our study suggests that if the Kolmogrov criterion for kinetic rates is
satisfied then one does not need a detailed non-equilibrium
calculation to obtain the stall force of multiple filaments/motors.
The same result can be obtained from a simple equilibrium statistical
mechanics calculation. Moreover, even if the Kolmogrov criterion for
kinetic rates is not satisfied for a system, our efficiency parameter $\alpha$ qualitatively predicts that the
cooperative effects in the stall force for multiple filaments (or motors)
is either enhanced or decreased as one increases the number of 
filaments (or motors). For a class of models discussed in this paper, we
clearly see a correlation between the numerical signs of $\alpha$ and
the force deviation $\Delta^{(2)}$. 
We would like to point out that the Kolmogorov or 
Wegschieder criterion for  detailed balance has been used earlier in the context of multiple growing filaments ~\cite{krawczykepl2011}  to describe the linear scaling of the stall force with the number of filaments.  
We, however, used the Kolmogorov criterion to systematically probe a range of models and derive the
conditions that must be obeyed by their respective kinetic rates, if the systems are
expected to be in ``equilibrium" at stall. More importantly, the conceptual advantage gained by couching this problem in the language of thermodynamic equilibrium is that, if any similar system is not in equilibrium at stall, 
then it is not at all guaranteed that the stall forces will be additive with the system size (number); in fact, non-additivity is the more likely outcome.



To illustrate the above point further, we take a concrete example of a two-state 
motor model from the existing literature~\cite{joanny2006prl,CasademuntPRL2009, oriola2013, oriola2014, julicher1997} (Fig.~\ref{fig:rachet}), which
demostrated the stall force non-additivity $f_s^{(N)} \neq N f_s^{(1)}$. Our analysis indicates the same conclusion without the need for any 
detailed simulation or theoretical calculations.  We identify that the violation of detailed balance in the transition rates between the states is a requirement for the spontaneous motion in this two-state 
Brownian ratchet model (see Fig.~\ref{fig:rachet}(a)). As a result, even for a single motor at stall, 
although the mean velocity flux of the motor is zero, the individual velocity fluxes in states-1 
and state-2 (with potentials ${\rm U}_1$ and ${\rm U}_2$, respectively) for the motor are not independently zero. 
In fact, only by observing the energy landscape one can say that, at stall, the mean particle flux in state-1 
and state-2  would be positive and negative, respectively. This is because,  the mean particle flux in state-2 is zero 
in the absence of any resisting force (flat potential, see Fig.\ref{fig:rachet}(a)) and would become negative with the application of a resisting force. To 
make the mean flux of the overall system zero, the effective mean flux in state-1 has to be positive under stall condition.
This non-equilibrium at stall for one motor manifests itself in the form of 
non-additivity of stall forces in the presence of multiple motors even though the only interaction between them is self-exclusion (Fig.\ref{fig:rachet}(a))~\cite{joanny2006prl, oriola2013, oriola2014}.  Interestingly, a discrete 
version of this two-state model~\cite{lacoste2007,lacoste2008}, shows 
 additivity of stall forces for multiple motors with only self-inclusion 
interactions (see Fig.~\ref{fig:rachet}(b)). This results from the fact that this discrete model for 
a single motor can easily be mapped to a biased random walk with only 
one track~\cite{fisher2001} -- the motors effectively  move on a tiled energy 
landscape, $\epsilon=\ln\frac{\omega_B\overrightarrow{\omega}_A}{\omega_A\overleftarrow{\omega}_B}$(see Fig.~\ref{fig:rachet}),
which, as argued in Section~II,  indicates that the motor can be interpreted to be in equilibrium at stall.

We can generalise the observations noted above and propose that the stall behaviour of passively interacting, processive motors greatly depends on the {\it topology} of the mechanochemical migration path of the motors.  Consequently, we generally propose that motor models analogous to the ones pioneered by Kolomeisky and Fisher~\cite{fisher1999, fisher2001} with a {\it single track} for motor migration will show additivity of stall force for multiple, passively interacting motors. On the other hand, the Brownian models with {\it multiple tracks}~\cite{julicher1997} demonstrably exhibit non-additivity of stall forces for multiple, even passively interacting motors~\cite{oriola2013, oriola2014}. To the best of our knowledge, the general conceptual framework that compares and contrasts the behaviour of these two major classes of molecular motor models, has not been provided before and is also one of the main contributions of this paper.

To summarize, collective force generation in biofilaments and molecular motors typically involve multi-step, complex internal processes and a variety of interactions between the individual entities as well as the source of the resisting force. In this paper we, however, have demonstrated that, even with very simple internal dynamics and also in the absence of any attractive or
repulsive interactions between individual components,  if a system of molecular motors or filaments is not in
equilibrium at stall we can expect non-additivity in their collective force generation.  To establish this result with reasonable certainty, we have analyzed, multiple seemingly disparate models, which nevertheless exhibit a common theme 
of non-equilibrium dynamics at stall leading to this cooperativity. The formalism developed in this paper should provide a general thermodynamics based framework with which to perform primary interpretation of experimental and theoretical results relating to collective force generation in biofilaments and molecular motors, before examining the system related specifics. 


\section*{Acknowledgments}
This work is supported by DST-Inspire research grant(T. B., IFA13 PH-64), CSIR India (Dipjyoti D., JRF award
no. 09/087(0572)/2009-EMR-I and Dibyendu D., no. 03(1326)/14/EMR-II. ), and DBT-IYBA (M.M.I., BT/06/IYBA/2012).

\section*{Appendix A}
\label{sec:Appendixa}
\section*{Simulation Method}
We have simulated our models using kinetic Monte-Carlo algorithm also known as the Gillespie algorithm,~\cite{BORTZ197510,Gillespiejpc1977,Gillespiejpc2007}, for the calculation of stall force of analytically unsolvable models .  Now we elaborate the exact method we have used to simulate ``generalized random hydrolysis model' for $n=3$ microtubule filaments.  We  start time evolution of the model system from a fixed initial length of $l_0=2000$ monomer at $t=0$ for all the microtubules. 
To obtain the time at which the next event will occur, we sum all possible event rates, $a_0$, and generate a random number $p$ between 0 and 1. 
Now $\tau~=~\frac{1}{a_0}~ ln \left( \frac{1}{p}\right)$ is the time of the next event.  To determine which event $k$ will occur at time $\tau$, we generate 
another random number $p_2$ between 0 and 1 and find $k$ which satisfies the condition $\sum_{i=1}^{k-1} a_i < a_0 p_2~<~\sum_{i=1}^k a_i$.  Now we 
repeat this processes till our system reaches a steady state. For $n=3$ microtubule, we have taken time to reach steady state as $t=1000 s^{-1}$.  Now at this 
point we reset our time to zero and start monitoring the position of largest microtubule filament with time. We plot this time evolution data and from 
the slope of the plot we get the velocity of the tip of largest filament. We perform this simulation for various forces and obtain corresponding velocity of the barrier. We then draw the force 
velocity curve.  The point at which this force velocity curve cuts the velocity axis is the stall  force.  To further verify the stall force, we calculate particle 
flux of the monomers for that force. For the calculation of flux, we just mark one filament and observe the number of monomers being added to that filament$(+1)$ and removed from that filament($-1$). Now we divide this count of monomers by the time window for which this bookkeeping has been done. If this flux is close to zero($>10^{-5}$) then we take that point as the stall force.

\section*{Appendix B}
\label{sec:Appendixb}
\section*{Connection between subunit flux and entropy production for multiple biofilaments at stall described using the random hydrolysis model of Section~\ref{sec:random}}

In the following text, we provide a simple, and perhaps crude, connection between the subunit flux $j_T(N)$ per filament, presented in Fig.~\ref{fig:2}c for the random hydrolysis model of Sec.~\ref{sec:random}, and the corresponding coarse grained entropy production $\dot{s}(N)$ per filament for $N$ filament at stall. 

Entropy production per filament is given by~\cite{lacoste2007,julicher1997}:
\bea
\label{eq:entropy}
\dot{s} = v f  + r\Delta \mu ,
\eea
where $f$, $v$, $\Delta \mu$ and $r$ are the external force, mean velocity, chemical potential for the subunit exchange and subunit exchange flux, respectively. At stall condition, $v = 0$, and hence, the entropy production is induced only by subunit exchange flux. For the simple T-D random hydrolysis model described in Sec.~\ref{sec:random}, the effective subunit fluxes (per filament) are $j_T$ and $j_D$, respectively. Moreover, at stall, since the total material influx, on an average, into the system of filaments is zero, $j_T = -j_D = j$. Since, in this model, the only subunit {\it on} rate is $U$ for the T subunits, the chemical potentials corresponding to T and D influx can simplistically be written as
\bea
\nonumber
\Delta \mu_T &=& \ln {U \over w_T}, \mbox{ and} \\ \label{eq:dmu}
\Delta \mu_D &=& \ln {U \over w_D}. 
\eea
Using Eqs.~\ref{eq:entropy}~and~\ref{eq:dmu}, the entropy production at stall condition can now be re-written as
\bea
\nonumber
\dot{s}(N) &=& j_T (N) \Delta \mu_T + j_D (N) \Delta \mu_D\\
&=& j(N) \ln{w_D \over w_T}.
\eea
Thus, a very simple argument proposes that entropy production per filament at stall condition is proportionate to the subunit flux per filament in the system of multiple filaments. 
\section*{References}
\bibliographystyle{jcp}
\bibliography{mybib}
\end{document}